\documentclass{article}

\usepackage{graphicx}
\usepackage{txfonts}
\usepackage{subcaption}

\usepackage{arxiv}

\usepackage[utf8]{inputenc} 
\usepackage[T1]{fontenc}    
\usepackage{hyperref}       
\usepackage{url}            
\usepackage{booktabs}       
\usepackage{amsfonts}       
\usepackage{nicefrac}       
\usepackage{microtype}      
\usepackage{lipsum}		
\usepackage{graphicx}
\usepackage{natbib}
\usepackage{doi}

\title{Cosmic Time Physics -- On the relation between cosmological redshift and fine structure constant variation}

\author{ \href{https://orcid.org/0000-0001-8318-6813}{\includegraphics[scale=0.06]{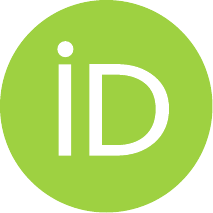}\hspace{1mm}Juan De Vicente}\thanks{Use footnote for providing further
		information about author (webpage, alternative
		address)---\emph{not} for acknowledging funding agencies.} \\
	Centro de Investigaciones Energ\'eticas, Medioambientales y Tecnol\'ogicas (CIEMAT),\\Avda. Complutense 40, E-28040, Madrid, Spain\\
              \texttt{juan.vicente@ciemat.es} }

\renewcommand{\shorttitle}{Cosmic Time Physics}

\hypersetup{
pdftitle={Cosmic Time Physics -- On the relation between cosmological redshift and fine structure constant variation},
pdfsubject={astrophysics,cosmology},
pdfauthor={Juan De Vicente-Albendea},
pdfkeywords={cosmology: observations -- cosmology: theory -- galaxies: distances and redshifts},
}

\begin{document}
\maketitle

\begin{abstract}
Almost a century ago, Hubble discovered the cosmological redshift of extragalactic objects. 
The Friedmann-Lema{\^\i}tre-Robertson-Walker (FLRW) metric was presented as a solution of Einstein's field equations for a homogeneous and isotropic universe. The metric includes a time-dependent factor $a(t)$, intended to explain the cosmological redshift. By contrast, for the Eintein's static universe ($a=1$), no reasonable redshift explanation was found. In this work, the Cosmic Time Physics (CTP) theoretical framework is developed. CTP moves the explanation of cosmological redshift from general relativity to electromagnetism domain. We show that the vacuum electric permittivity $\epsilon_0$ and the vacuum magnetic permeability $\mu_0$ can vary inversely one each other over cosmic time, maintaining the speed of light $c$ constant, while conducting the change on the vacuum impedance $Z_0$ and on the fine structure constant $\alpha$. This variation downscales the atomic energy levels with cosmic backtime, redshifting the wavelength and frequency exactly in the same manner they are observed, while maintaining the atomic quantification relations. Note that the increase on $\alpha$ with cosmic time has gone unnoticed experimentally so far since the search is performed on rest-frame (de-redshiftted signals), in spite of the manifestation of such variation is precisely the redshift. The application of  CTP to general relativity drive to an angular-redshift relation $d_A(z)$ as a function of the age of the universe $t_0$ and its curvature $R_0$. As a first approximation, we show that CTP $d_A(z)$ is able to reproduce the LCDM $d_A(z)$ curve with $R_0=1800$ Mpc and $t_0=15.57$ Gly. Finally, the Friedmann equations without scale factor ($a=1$) are used to derive the requirements for the stability of CTP universe. 
\end{abstract}


\section{Introduction}
During the 20th century the foundations of modern cosmology were established. The field equations of general relativity were formulated by \cite{einstein1915feldgleichungen}. The definition of new metrics based on the cosmological principle with the properties of homogeneity and isotropy allowed the physicists the application of Einstein's field equations to the universe. While Einstein defined a static metric, \cite{friedmann1922krummung} deduced mathematically a non-stationary model with a time-dependent factor $a(t)$. The solution was independently derived by  \cite{lemaitre1927lemaitre}, interpreting a(t) as a scale factor of an expanding universe. The work was completed by \cite{robertson1933relativistic} and \cite{walker1937milne} in what is known as the \textit{Friedmann-Lema{\^\i}tre-Robertson-Walker} (FLRW) metric.  
	
	Contemporaneously to these achievements, a correlation between redshifts and distances for extragalactic sources was found by  \cite{hubble1929relation}. The origin of this correlation was subject of intense debate between proponents of static and expanding universes on the 1930s (\cite{kragh2017universe}). The fault of the Einstein's static universe to explain the redshift of galaxies leaned the balance to the FLRW metric, whose time dependent factor $a(t)$ can explain the cosmological redshift (\cite{tolman1934relativity}). The FLRW model describes a solutions to the Einstein's field  equations for a homogeneous and isotropic universe. The evolution and fate of the Universe depends on the nature of different density components, i.e., radiation, matter, curvature and dark energy. 
    
	Different cosmological tests were proposed to probe whether the Universe is expanding or remains static. \cite{tolman1930estimation} predicted that in an expanding universe, the surface brightness of a receding source with redshift $z$ will be dimmed by $\sim(1+z)^{-4}$. Consequently to Tolman's prediction, the equation $d_L=d_A(1+z)^\gamma$, with $\gamma=2$ was established between \textit{luminosity distance} $d_L$ and \textit{angular diameter distance} $d_A$ for a expanding universe. There are contradictory results on this kind of test, some of them claim for expansion (\cite{lubin2001tolman,sandage2010tolman}) while others (\cite{lerner2006evidence,crawford2010observational,brynjolfsson2006surface}) are advocating for a static universe. Nevertheless, neither expanding nor static studies are conclusive, since they depend on an uncertain possible galaxy evolution. In this work, we develop the Cosmic Time Physics theory that offer an alternative explanation to cosmological redshift, conducting to a new consistent cosmological model.

 The rest of the paper is organized as follows: 
 Section~\ref{sec:stdModel} describes the foundations of the expanding universe. Section~\ref{sec:CTP} present the Cosmic Time Physics theoretical framework. The clarification of some issues related to the CTP theory are exposed in Section~\ref{sec:Discussion}.  The conclusions are presented in Section~\ref{sec:conclusions}.

\section{Foundations of the expanding universe}
\label{sec:stdModel}

The expanding universe rest on the Einstein's field equation given by

\begin{eqnarray}
\label{eq:EFE}
	R_{\mu\nu}-\frac{1}{2}g_{\mu\nu}R+\Lambda g_{\mu\nu}=\frac{8 \pi G}{c^4} T_{\mu\nu}
\end{eqnarray}

\noindent
where $R_{\mu\nu}$ is the Ricci curvature tensor, ${\displaystyle R}$ is the scalar curvature, $T_{\mu\nu}$ is the energy momentum tensor and  $g_{\mu\nu}$ metric tensor. The form of $g_{\mu\nu}$ for a homogeneous and isotropic universe is known as the FLRW metric and is given by

\begin{eqnarray}
  \label{eq:flrw0}
   -c^2d\tau^2=-c^2dt^2+a(t)^2\left[\frac{dr^2}{1-kr^2}+r^2d\Omega^2\right]
\end{eqnarray}

\noindent
being

\begin{eqnarray}
  \label{eq:flrw_angles}
   d\Omega^2=d\theta^2+sin^2\theta d\phi^2
\end{eqnarray}

\noindent where k describes the curvature, while a(t) is a time-dependent factor commonly interpreted as the scale factor of an expanding universe. Note that a(t)=1 corresponds to the Einstein's static universe. On the other hand, it can be demonstrated that $a(t)$ meets the Friedmann equations


\begin{equation}
	\label{eq:friedmann1}
	H^2 = \frac{8\pi G}{3} \rho - \frac{k }{a^2}
\end{equation}


\begin{equation}
	\label{eq:friedmann2}
	\frac{\ddot{a}}{a} = -\frac{4\pi G}{3} ( \rho + 3 p )
\end{equation}

where $H$ is the Hubble parameter, $a(t)$ the scale factor, $\rho$ the density, $p$ the pressure and $k$ the curvature with values $k=0,1,-1$ for flat, closed and open universes respectively. \\

There are different distance ladders relating theory and observations. Let us to provide a brief summary of some distance definitions and their relations with normalized densities ($\Omega_M$, $\Omega_r$, $\Omega_\Lambda$, $\Omega_k$), corresponding to matter, radiation, cosmological constant and curvature (\cite{hogg1999distance}) respectively. The first Friedmann equation can be expressed from the Hubble parameter $H$ at any time, and the Hubble constant $H_0$ today as

\begin{eqnarray}
  \label{eq:friedmann}
   \frac{\dot{a}(t)^2}{a(t)^2}=H^2=H_0^2 E(z)^2
\end{eqnarray}

\noindent where 

\begin{eqnarray}
\label{eq:ez}
  E(z)=\sqrt{\Omega_K(1+z)^2+\Omega_\Lambda+\Omega_M(1+z)^3+\Omega_r(1+z)^4}
\end{eqnarray}

\noindent By integrating Eq. ~\ref{eq:flrw0} along with Eq. ~\ref{eq:friedmann} one can obtain the line of sight \textit{comoving distance} $d_C$ as

\begin{eqnarray}
\label{eq:comovingDistance}
  d_C=d_H\int_{0}^{z}\frac{dz'}{E(z')}
\end{eqnarray}

\noindent where $d_H=c/H_0=3000 h^{-1} Mpc$ is the \textit{Hubble distance}. From the same equations one can get the \textit{transverse comoving distance} $d_M$ as

\begin{eqnarray}
\label{eq:transverseComovingDistance}
d_M= 
\left \{ 
 \begin{array}{ccc}
    d_H\frac{1}{\sqrt\Omega_k}\sinh[\sqrt\Omega_kd_C/d_H]&for&\Omega_k>0 \\ 
    d_c&for&\Omega_k=0 \\
    d_H\frac{1}{\sqrt{|\Omega_k|}}\sin[\sqrt{|\Omega_k|}d_C/d_H]&for&\Omega_k<0 \\
    \end{array}
\right \}
\end{eqnarray}

\noindent
With respect to observable quantities, the \textit{angular diameter distance} $d_A$ is defined as the ratio between the object physical size $S$ and its angular size $\theta$ 
 
\begin{eqnarray}
\label{eq:angularDiameterDistanceTh}
  d_A=\frac{S}{\theta}
\end{eqnarray}

\noindent
The \textit{angular diameter distance} is related to the \textit{transverse comoving distance} by
 
\begin{eqnarray}
\label{eq:dAdM}
  d_M=d_A(1+z)
\end{eqnarray}

\noindent
where z is the redshift. On the other hand, the \textit{luminosity distance} defines the relation between the bolometric flux energy $f$ received at earth from an object, to its bolometric luminosity L by means of

\begin{eqnarray}
\label{eq:fluxEnergy}
 f= \frac{L}{4\pi d_L^2} 
\end{eqnarray}

\noindent
or finding $d_L$

\begin{eqnarray}
\label{eq:DL_th}
 d_L=\sqrt{\frac{L}{4\pi f}} 
\end{eqnarray}

\noindent
The relation between $d_L$ and $d_M$ is given by

\begin{eqnarray}
\label{eq:luminosity_vs_transverseComoving}
d_L= d_M(1+z)
\end{eqnarray}

\noindent
and taking into account Eq. ~\ref{eq:dAdM}

\begin{eqnarray}
\label{eq:luminosity2_vs_angular2}
d_L^2= d_A^2(1+z)^4
\end{eqnarray}

\noindent
There are four (1+z) factors affecting to flux energy diminution. Two come from the elongation of the initial distance $d_A$ by a factor of $(1+z)$ due to universe expansion according to the inverse square law. Another factor comes from the time dilation due to universe expansion that reduces the photon emission/reception rate by $(1+z)^{-1}$. The last factor comes from the cosmological wavelength redshift that decrease the energy of photons by $(1+z)^{-1}$. Therefore, a relevant relation is established between the \textit{angular diameter distance} and the \textit{luminosity distance} in the \textit{expanding universe} as

\begin{eqnarray}
\label{eq:stDistancesRelation}
d_L=d_A(1+z)^2               \qquad (expanding\ universe)
\end{eqnarray}

\noindent
Eq. ~\ref{eq:stDistancesRelation} is commonly known as Etherington distance-duality relation. It is important to clarify that the original \cite{etherington1933lx} equation is the reciprocity theorem for a local (i.e. non-expanding) universe:

\begin{eqnarray}
\label{eq:neDistancesRelation}
d_L=d_A(1+z)               \qquad (non-expanding\ universe)
\end{eqnarray}

It was reformulated for an expanding universe by adding another $(1+z)$ factor.

\section{Cosmic Time Physics}
\label{sec:CTP}
In this section, we develop the Cosmic Time Physics (CTP)
theoretical framework, where the cosmological redshift can be explained by the integration of cosmic time on
physical equations. As a reward, CTP also describes the long term
behaviour of some fundamental physical laws. The hints came from the scrutiny of the invariability of some physical constants. While some constants have a profound meaning in the physical theories, other simply
have shown its invariability in a tiny period of time compared with cosmic time. According to \cite{uzan2011varying} review, some constants have a much deeper role in physics than others. \cite{levy1979importance} classify the constants in different categories: A, B, C. Class A corresponds to constants of a particular system, class B integrates constants characteristic of a physical phenomena, while C corresponds to the class of universal constants, the most fundamental constants where $h$ and $c$ are classified. On the other hand, \cite{dirac1937cosmological} suggests that physical dimensionless constants should be variables depending on the cosmic time. Among these dimensionless parameters we have the fine structure constant $\alpha=1/137$ and the ratio $m_p/m_e=1836$. In this section, we explore the possibility of the variation of $\alpha$ with cosmic time.

\subsection{CTP -- On the relation of the cosmological redshift and the fine structure constant}
\label{sec:alpha}

The cosmological redshift is a phenomenon observed on the light coming from extragalactic objects. We know about it since the spectral lines emitted by those atoms appear shifted to the red with respect to what we observe at the laboratory. To analyse the phenomenon, we can go through the equations that govern the gross emission lines of the hydrogen atom as a function of the fine structure constant $\alpha$. $\alpha$ is a dimensionless amount that measure the strength of electromagnetic interactions. It is assumed to be invariable with time since it is defined as a function of other fundamental physical constants.

The energy levels $E_n$, the orbital radius $r_n$ and the velocity of the electron $v_n$ of the hydrogen atom as a function of $\alpha$ are given by

\begin{eqnarray}
\label{eq:energy}
  E_n = -m_e c^2 \alpha^2\frac{1}{2n^2}
\end{eqnarray}
 
\begin{eqnarray}
\label{eq:radius}
   r_n = \frac{n^2 \hbar}{m_e c \alpha}n^2            
\end{eqnarray}

\begin{eqnarray}
\label{eq:velocity}
  v_n = \frac{\alpha c}{n}      
\end{eqnarray}

\noindent And the energy $E_{n,m}$, frequencies $\nu_{n,m}$ and wavelengths $\lambda_{n,m}$ of the emission lines corresponding to fundamental transitions in the hydrogen atom can be expressed as

\begin{eqnarray}
\label{eq:transitionEnergy1}
 E_{n,m} = m_e c^2 \alpha^2 {\left(\frac{1}{m^2} - \frac{1}{n^2}\right)}
\end{eqnarray}

\begin{eqnarray}
\label{eq:frequency1}
  \nu_{n,m} = \frac{m_e c^2 \alpha^2}{4\pi\hbar} \left(\frac{1}{m^2} - \frac{1}{n^2}\right)
\end{eqnarray}

\begin{eqnarray}
\label{eq:wavelength1}
  \lambda_{n,m} = \frac{4\pi\hbar}{m_e c \alpha^2} \left(\frac{1}{m^2} - \frac{1}{n^2}\right)^{-1}
\end{eqnarray}

At this point, we rise the reasonable hypothesis that the extragalactic atoms at redshift $z$ emit the photons with the energy ($E_z$), frequency ($\nu_z$) and wavelength ($\lambda_Z$) we observe now. Therefore 

\begin{eqnarray}
\label{eq:transitionEnergy2}
 E_z\propto \frac{1}{1+z}
\end{eqnarray}

\begin{eqnarray}
\label{eq:frequency2}
  \nu_z \propto \frac{1}{1+z} 
\end{eqnarray}

\begin{eqnarray}
\label{eq:wavelength2}
  \lambda_z \propto (1+z) 
\end{eqnarray}

Leaving apart the different values of n and m, we can see from Eq.~\ref{eq:transitionEnergy1}-~\ref{eq:frequency1} that $E_z$, $\nu_z$ and $\lambda_z$ depend on the physical constants: $h$, $c$, $m_e$, $\alpha$. Thus, to have the observed values of $E_z$, $\nu_z$ and $\lambda_z$, we have to allow one of the physical constants to vary with redshift. Following the suggestion of \cite{dirac1937cosmological}, who claimed that the dimensionless constants should be variables depending on cosmic time, we explore the possibility of the variation of $\alpha$ with redshift. Taking into account Eq.~\ref{eq:transitionEnergy2}-~\ref{eq:wavelength2}, Eq.~\ref{eq:transitionEnergy1}-~\ref{eq:wavelength1} becomes  

\begin{eqnarray}
\label{eq:transitionEnergy3}
 E_z \propto \alpha_z^2 \propto \frac{1}{1+z}
\end{eqnarray}

\begin{eqnarray}
\label{eq:frequency3}
  \nu_z \propto \alpha_z^2 \propto \frac{1}{1+z} 
\end{eqnarray}

\begin{eqnarray}
\label{eq:wavelength3}
  \lambda_z \propto \frac{1}{\alpha_z^2} \propto (1+z) 
\end{eqnarray}

To sum up,

\begin{eqnarray}
\label{eq:all2}
  \alpha_z^2 \propto E_z \propto \nu_z \propto \frac{1}{\lambda_z} \propto \frac{1}{1+z}
\end{eqnarray}   

On the other hand, $\alpha$ is defined as

\begin{eqnarray}
\label{eq:alpha1}
	\alpha=\frac{e^2}{2hc\epsilon_0}               
\end{eqnarray}

\noindent which can also be expressed in the form

\begin{eqnarray}
\label{eq:alpha2}
	\alpha=\frac{e^2}{2h}Z_0               
\end{eqnarray}

\noindent where $Z_0$ is the vacuum impedance given by

\begin{eqnarray}
\label{eq:impedance}
	Z_0=\sqrt{\frac{\mu_0}{\epsilon_0}}               
\end{eqnarray}

Therefore, according to Eq.~\ref{eq:all2} and Eq.~\ref{eq:alpha2}, and considering $h$ and $e$ more fundamental constants, the cause of CTP cosmological redshift points to $Z_0$. But as we have assume the speed of light $c$ as a fundamental constant, it is required 

  \begin{eqnarray}
\label{eq:mu_z}
   \mu_z \propto \frac{1}{\epsilon_z}
\end{eqnarray}

\noindent where $\epsilon_z$ and $\mu_z$ accounts for the electric permittivity and magnetic permeability of vacuum at cosmological redshift z, meeting also 

\begin{eqnarray}
\label{eq:c_prop_1}
   c^2 \propto \frac{1}{\mu_z\epsilon_z} \propto 1
\end{eqnarray} 

\noindent i.e. the speed of light $c$ remains constant as required.

To sum up, we have the following relations

\begin{eqnarray}
\label{eq:all3}
  \alpha_z^2 \propto Z_z^2 \propto \frac{\mu_z}{\epsilon_z} \propto \mu_z^2 \propto \frac{1}{1+z}
\end{eqnarray} 

\noindent where $Z_z$ is the vacuum impedance at cosmological redshift z.

To related the above equations with cosmic time, we can define within CTP the age of the universe at redshift z as

\begin{eqnarray}
\label{eq:tz}
  t_z=\frac{t_0}{1+z}
\end{eqnarray} 		

\noindent being $t_0$ the current age. 

In that case, Eq.~\ref{eq:all2} and Eq.~\ref{eq:all3} can be re-written as a function of cosmic time as

\begin{eqnarray}
\label{eq:all4}
  \frac{t_z}{t_0}=\frac{\alpha^2}{\alpha_0^2}=\frac{Z_z^2}{Z_0^2}=\frac{\mu^2}{\mu_0^2}=\frac{E_z}{E_0}=\frac{\nu_z}{\nu_0}=\frac{\lambda_0}{\lambda_z}=\frac{1}{1+z}
\end{eqnarray} 		

\noindent And if we define the relative cosmic time at redshift $z$ as

\begin{eqnarray}
\label{eq:tauz}
  \tau_z=\frac{t_z}{t_0}
\end{eqnarray}	

\noindent we have

\begin{eqnarray}
\label{eq:Ez_vs_tauz}
  E_z=E_0 \tau_z 
\end{eqnarray}	

\noindent which shows that the energy levels of the hydrogen atom grows slowly with the relative cosmic time.

To summarize, the standard explanation of cosmological redshift assumes that some property of space (volume) is changing over cosmic time. According to Eq.~\ref{eq:all4}, CTP explains the cosmological redshift as changes over time of another property of space as the fine structure constant or the impedance of vacuum. Note that there is a key difference between the vacuum impedance and a medium impedance. In the last case, the wavelength adapts to the wave velocity decrease, while the frequency remains unchanged. Since velocity of the wave in the vacuum is the constant speed of light $c$, both frequency $\nu_z$ and wavelength $\lambda_z$ of photons do not change from emission to reception --in spite of vacuum impedance variation--, as otherwise observed.

\subsection{CTP -- Redshift, time and distance}
\label{sec:redshift_time_distance}
The cosmological redshift in LCDM is explained as the effect of the universe expansion. Contrary, the cosmological redshift in CTP is explained from the electromagnetism. Therefore, the relation of the cosmological redshift with time and space is not exactly the same in both theories. Let us to explain in this section how the cosmological redshift is related to time and distance in CTP.

The cosmological redshift is defined in both theories as
\begin{eqnarray}
\label{eq:redshift}
   z=(\lambda_z-\lambda_0)/\lambda_0
\end{eqnarray} 	
	
\noindent where $\lambda_z$ is the wavelength, measured at earth, corresponding to a photon emitted from an event on a distant source. Meanwhile $\lambda_0$ is the wavelength that we would measure at the laboratory from a similar event.

Let us to focus on Eq.~\ref{eq:tz}. The absolute cosmic time $t_0$ can be expressed as  

\begin{eqnarray}
\label{eq:abstime}
   t_0=t_z+\bar{t_z}
\end{eqnarray}

\noindent where $\bar{t_z}$ is the lookback time of the event at redshift z. Substituting Eq.~\ref{eq:abstime} in Eq.~\ref{eq:tz}, and solving for z we have

\begin{eqnarray}
\label{eq:redshift2}
   z=\frac{\bar{t_z}}{t_z}
\end{eqnarray}

\noindent We can also express the lookback time $\bar{t_z}$ as a function of $t_0$ and $z$ as 

\begin{eqnarray}
\label{eq:lookbacktime}
   \bar{t_z}=t_0 \frac{z}{1+z} 
\end{eqnarray}

Regarding the distance, we can define the lookback distance $\bar{d_z}$ as the distance to a cosmological event produced at $\bar{t_z}$ as
 
\begin{eqnarray}
\label{eq:dz}
   \bar{d_z}=c \bar{t_z}
\end{eqnarray}

\noindent or as a function of $z$ as

\begin{eqnarray}
\label{eq:dz}
   \bar{d_z}=c t_0 \frac{z}{1+z}
\end{eqnarray}

\noindent that gives the distance to the event as a function of redshift and the age of the universe.

\subsection{CTP -- Conditions for the stability}
\label{sec:stability}
In this section, we derive the conditions for the stability of CTP universe described above. Let us set a(t)=1 in Friedmann equations given by Eq.~\ref{eq:friedmann1} and Eq.~\ref{eq:friedmann2}. The resulting equations are

\begin{eqnarray}
	\label{eq:friedmann11}
	 8\pi G \rho=3k
\end{eqnarray}

\begin{eqnarray}
	\label{eq:friedmann12}
	 4\pi G (\rho + 3p)=0
\end{eqnarray}

\noindent Deriving both equations with respect to time and substituting one in the other we have

\begin{eqnarray}
	\label{eq:friedmann13}
	\dot{\rho}=0
\end{eqnarray}

\begin{eqnarray}
	\label{eq:friedmann14}
	\dot{p}=0
\end{eqnarray}

\noindent Therefore

\begin{eqnarray}
	\label{eq:friedmann15}
	\rho=const
\end{eqnarray}

\begin{eqnarray}
	\label{eq:friedmann16}
	p=const
\end{eqnarray}

\noindent Expressing $\rho$ and $p$ in their components

\begin{eqnarray}
	\label{eq:friedmann14}
	\rho_m+\rho_{r}+\rho_{vac}=const
\end{eqnarray}

\noindent where $\rho_m$, $\rho_r$ and $\rho_{vac}$ are the matter density, the radiation density and vacuum energy density respectively. While the pressure equation becomes 

\begin{eqnarray}
	\label{eq:friedmann15}
	p_{rad}+p_{vac}+p_m=const
\end{eqnarray}

\noindent where
$p_r, p_{vac}$ are the radiation and vacuum pressures, and $p_m$ the matter pressure that is zero for non-relativistic matter.

Therefore, the global sum of densities on one hand, and the global sum of pressures on the other, should remain constant. For the universe to be stable, it is required that the different components of $\rho$ and $p$ depend on time in such a way that any variation in one component can be absorbed by a variation in another one. In addition, note that since in CTP model there is no expansion (i.e. the volume is constant), Eq.~\ref{eq:friedmann14} can be transformed in

\begin{eqnarray}
	\label{eq:friedmann16}
	 E_m(t)+E_r(t)+E_{vac}(t)=E_u
\end{eqnarray}
 
\noindent where $E_m(t)$ is the matter energy, $E_r(t)$ the radiation energy, $E_{vac}(t)$ the vacuum energy and $E_u$ the universe energy. Therefore, the energy of the universe $E_u$ would be constant, allowing the exchange of energy among the different components: matter, radiation and vacuum. 

\subsection{CTP -- Angular distance} 
In CTP theory, the cosmological redshift is not explained by the universe expansion, thus the transverse comoving distance $d_M$ defined in Eq.~\ref{eq:transverseComovingDistance} loses its meaning. Nevertheless, its role relating distances and curvature is taking by the angular distance $d_A$ taking the form

\begin{eqnarray}
\label{eq:CTP_angular_distance}
d_A= 
\left \{ 
 \begin{array}{ccc}
    R_0 sinh[\bar{d_z}/R_0&for&k=-1 \\ 
    \bar{d_z}&for&k=0 \\
    R_0 sin[\bar{d_z}/R_0&for&k=1 \\ 
    \end{array}
\right \}
\end{eqnarray}

\noindent where $R_0$ is the radius of the universe for the curvature cases (k=-1, k=1).
 
\begin{figure} 
  \centering
         \includegraphics[width=0.45\textwidth]{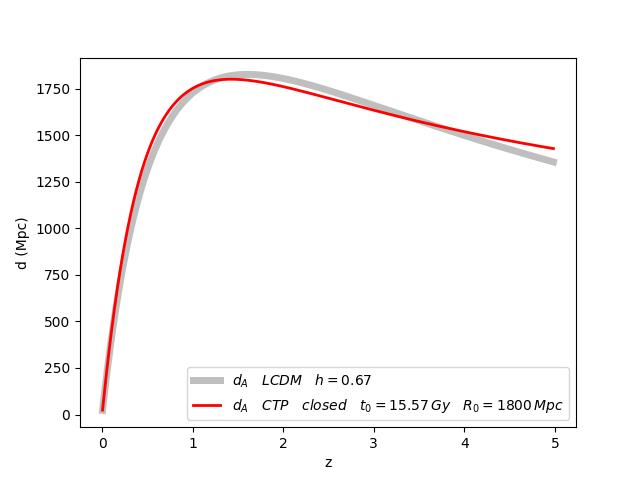}
	\caption{Fit of CTP angular distance to LCDM angular distance (assumed as representative of data) with the two free parameters: the radius ($R_0$) and age of the universe ($t_0$).}
 \label{fig:CTP_angular_distance}
\end{figure}

To evaluate whether CTP is able to explain observational data, we have taken as reference the angular distance $d_A$ of LCDM. The $d_A$ of LCDM is assumed to represent properly the angular distance measurements of CMB (Cosmic Microwave Background) and BAO (Baryon Acoustic Oscillations) data. Then, we have tried to fit $d_A$ predicted by CTP open, flat and closed universes to LCDM $d_A$. While CTP open and flat fit attempts were unsuccessful, CTP closed $d_A$ provides a reasonable  and encouraging good fitting. Thus, Fig.~\ref{fig:CTP_angular_distance} shows the fit of angular distance $d_A$ for CTP closed universe to the LCDM $d_A$. The best fit corresponds to a closed universe of radius $1800$ Mpc and $15.57$ Gly. Note that it is only a first approximation. The precise values and uncertainties will be obtained by the fit to angular distance data. 

Note that if the universe is closed, i.e. $k=1$, according to Eq.~\ref{eq:friedmann1} and Eq.~\ref{eq:friedmann16}, the energy of the universe is constant and greater than zero

\begin{eqnarray}
 \label{eq:Epositiva}
	  E_u>0
\end{eqnarray}

\section{Discussion}
\label{sec:Discussion}

There are some issues related to CTP results that require clarification. One is related to the lack of detection of changes in the fine structure constant value through different experiments. It is easily explained, since the search is performed on rest-frame (i.e. de-redshifted signals), while the signal of the change searched is precisely, and unexpectedly, the redshift. 

Eq.~\ref{eq:friedmann16} and Eq.~\ref{eq:Epositiva} derived from Friedmann equations, only establish that the global energy is a positive constant, but do not describe how their components changes over time. In the case of the atomic energy levels, Eq.~\ref{eq:Ez_vs_tauz} shows that they grow with cosmic time, thus an equivalent decrease should be expected in its corresponding vacuum energy. It can be explained in terms of the variation of the fine structure constant. As the vacuum energy decreases, the lower both the vacuum fluctuations and the shield they produce on the electromagnetic strength interaction, i.e. fine structure constant grows. 

An interesting fact, it is that the radiation coming from background galaxies is on average of lower energy than the foreground one, reducing the probability of interaction with intermediate matter, and hence lowering its illumination. 

Another remarkable fact is that the variation of the fine structure constant with cosmic time not only affects to atomic energy levels, but to proton-proton interaction, and hereby it may affects to star formation and Chandrasekhar limit, responsible for the supernovae Type-Ia explosions. Thus, the cosmology based on standard candles should be revised in the context of CTP. 

CTP is a theory under construction, thus we should not expect a promptly TCP response to all physical phenomena. In this work, we have laid the first stone. The atomic levels dependence with cosmic time not only affects to cosmology but also to other sciences. If CTP is confirmed, it is expected a cascade and coordinated development of CT-theorical physics, CT-particle physics, CT-nucleosynthesis, CT-chemistry, CT-astrophysics and CT-biology among others. 

Finally, let us to reveal the illusion of the expansing universe corresponding to the FLRW metric. Dividing the FLRW metric (Eq.~\ref{eq:flrw0}) by $a(t)^2$ we have 

\begin{eqnarray}
  \label{eq:flrw1}
   -c^2\frac{d\tau^2}{a(t)^2} =-c^2\frac{dt^2}{a(t)^2}+\left[\frac{dr^2}{1-kr^2}+r^2d\Omega^2\right]
\end{eqnarray}

Now, making the change of variables, $dt'=dt/a(t)$ and $d\tau'=d\tau/a(t)$, the Eq.~\ref{eq:flrw1} can be expressed as

\begin{eqnarray}
  \label{eq:flrw2}
   -c^2 d\tau'^2 =-c^2 dt'^2+\left[\frac{dr^2}{1-kr^2}+r^2d\Omega^2\right]
\end{eqnarray}

\noindent where the scale factor $a(t)$ disappear, fading the expanding universe illusion. Note, that the FLRW is still valid for a homogeneous and isotropic non-expanding universe. 

\section{Conclusion}
\label{sec:conclusions}

In the 1930s, a debate was held on different possible explanations of the cosmological redshift discovered by Hubble. The FLRW metric for a homogeneous and isotropic universe emerged as a solution of Einstein's field equations. The metric included a time dependent factor $a(t)$, that was interpreted as a scale factor of an expanding universe, able to explain the cosmological redshift. By contrast, no reasonable explanation was found for the cosmological redshift in the Einstein's static universe (a=1).

In this work, CTP theory is presented. Following the expectations expressed by Dirac in 1937, regarding the variation with cosmic time of some dimensionless physical constants, we show that the increase of fine structure constant --and hence the vacuum impedance-- with cosmic time, would be able to explain the cosmological redshift, leaving other more fundamental constants as the speed of light $c$ or the Plank constant $h$ untouched, unlike the attempts performed by \cite{dicke1957gravitation}.

Based on the new relations of redshift with time and distance derived from CTP, we construct the equations that describe the angular distance $d_A$ for a flat, closed and open universes. It is shown that CTP $d_A$ for a closed universe can roughly mimic LCDM $d_A$, providing a feasible alternative to cosmological redshift explanation. The application of CTP constraints to Friedmann equations, drives to a closed universe with global constant energy $E_u>0$, distributed among variable matter, radiation and vacuum energies, which guarantee its stability.

CTP is a novel theory that requires further development and the validation with cosmological data. Finally, note that, whenever confirmed, CTP opens a new insight in physics that would affect to many branches of science.



\section*{Acknowledgements}
Funding support for this work was provided by the Autonomous Community 
of Madrid through the project TEC-2024/TEC-182 MAD4SPACE.


\bibliographystyle{unsrtnat}
\bibliography{ctp}  

\end{document}